\documentclass[runningheads,envcountsame]{llncs}

\usepackage{amsfonts}
\usepackage{amssymb}
\usepackage{amsmath}
\usepackage{mathtools}
\usepackage{fixltx2e}
\usepackage[ruled,noend]{algorithm2e}
\usepackage[noend]{algpseudocode}
\usepackage{mathpartir}
\usepackage{stmaryrd}
\usepackage{amsfonts}
\usepackage{bbold}
\usepackage{marvosym}
\usepackage{scalerel}
\usepackage{todonotes}
\usepackage{multirow,bigdelim}
\usepackage{placeins}
\usepackage{array}
\usepackage{enumitem}
\usepackage{caption}
\usepackage{subcaption}
\usepackage{pgfplots}
\usepackage{placeins}

\LinesNumbered

\makeatletter
\RequirePackage[bookmarks,unicode,colorlinks=true]{hyperref}%
   \def\@citecolor{blue}%
   \def\@urlcolor{blue}%
   \def\@linkcolor{blue}%

\def\orcidID#1{\smash{\href{http://orcid.org/#1}{\protect\raisebox{-1.25pt}{\protect\includegraphics{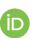}}}}}
\makeatother

\usepackage{cleveref}

\makeatletter
\newcommand{\nosemic}{\renewcommand{\@endalgocfline}{\relax}}
\newcommand{\dosemic}{\renewcommand{\@endalgocfline}{\algocf@endline}}

\makeatother

\newenvironment{subroutine}[1][!htb]{%
    \begin{algorithm}[#1]%
}{\end{algorithm}}

\crefname{subroutine}{Subroutine}{Subroutines}
\Crefname{subroutine}{Subroutine}{Subroutines}

\newcommand{\N}{\mathbb{N}}

\newcommand{\lstar}{\ensuremath{\texttt{L}^{\!\star}}}
\newcommand{\kv}{Kearns--Vazirani}

\newcommand{\eset}{\emptyset}
\newcommand{\eword}{\varepsilon}

\newcommand{\lang}{\mathcal{L}}

\newcommand{\aut}{\mathcal{A}}
\newcommand{\reach}{\mathsf{reach}}

\newcommand{\mq}{\mathsf{mq}}
\newcommand{\eq}{\mathsf{eq}}

\SetKwComment{Comment}{/* }{ */}

\begin{document}
\title{Tree-Based Adaptive Model Learning\thanks{Full implementation and experiment results available at \url{https://github.com/UCL-PPLV/learnlib}.}}

\titlerunning{Keep Learning}
\author{Tiago Ferreira\inst{1}\orcidID{0000-0002-6942-0228} \and
Gerco van Heerdt\inst{1}\orcidID{0000-0003-0669-6865} \and
Alexandra Silva\inst{2,1}\orcidID{0000-0001-5014-9784}}
\institute{University College London \\ \email{\{t.ferreira,gerco.heerdt\}@ucl.ac.uk} \and
Cornell University \\ \email{alexandra.silva@cornell.edu}}
\maketitle
\emph{A lot of the work the authors did in the last years on learning can be rooted back to learning from Frits. Gerco and Alexandra spent a few years in Nijmegen where Frits’ enthusiasm for Angluin’s algorithm infected them and made them want to keep learning. Later on, Tiago first learned about automata watching Frits' model learning video for CACM. On this landmark birthday, we thank Frits for his inspiration and wish him many happy returns!}
\begin{abstract}
    We extend the \kv\ learning algorithm to be able to handle systems that change over time.
    We present a new learning algorithm that can reuse and update previously learned behavior, implement it in the LearnLib library, and evaluate it on large examples, to which we make small adjustments  between two runs of the algorithm.
    In these experiments our algorithm significantly outperforms both the classic \kv\ learning algorithm and the current state-of-the-art adaptive algorithm.
\end{abstract}
\section{Introduction}

Formal methods have a long tradition and have had much success in critical applications, e.g. in the medical, space, and hardware industries. The last decade saw a rise in the use of formal methods in the software industry, with dedicated large teams in many companies, notably AWS and Facebook. This has caused a shift in focus to develop techniques that are helpful towards catching the most bugs as quickly as possible instead of performing complete verification~\cite{bornholt2021}.

The use of models to analyze system specifications, often pre-production, is common in certain domains, but requires expert knowledge to design the model and, throughout the system's life, update it. Motivated by the difficulties of building models, an automated technique called model learning~\cite{frits-cacm} has been developed and used in the analysis of a range of (black-box system) implementations. One particularly successful application is that of network protocol implementations, e.g. TCP~\cite{Fiterau-Brostean16}, SSH~\cite{Fiterau-Brostean17}, and QUIC~\cite{FerreiraBD021}.

Classic active model learning algorithms like \lstar~\cite{angluin1987}, \kv~\cite{kv1994}, and TTT~\cite{ttt2014} exist for a number of automata types (deterministic, input-output, weighted, register) and have enabled analysis of numerous systems by providing faithful models to be used in model checking.
However, these algorithms suffer from a common issue: systems often change faster than we can learn their models, as there is an inherent assumption that the learning process is running from scratch every time.
As such, keeping a model up-to-date becomes quite challenging, often needing manual intervention from an expert.
This does however introduce the chance of producing a model that is not actually correct, while being an extremely laborious task for any moderately sized complex system.

In this paper, we present an \emph{incremental} model learning algorithm that can cope with evolving systems more effectively, and does not have to restart the learning process when a change occurs, providing a gain in efficiency when learning systems that undergo changes at known moments in time.


The work in the present paper can be seen as advancing the state of the art in {\em adaptive automata learning}~\cite{groce2002,chaki2008,windmuller2013,huistra2018,damasceno2019}.
Whereas previous work adapted the \lstar{} algorithm to reuse part of a data structure from a previous run, our adaptive algorithm is the first to use the more efficient Kearns--Vazirani algorithm.
Although we focus on developing an algorithm for deterministic finite automata (DFAs), the techniques we developed can be transferred to other automata models.
We expect this work and subsequent developments on adaptive learning will bring us closer to meet the specific needs of employing formal methods in fast moving environments commonly seen in the industry, and in large evolving systems.

The paper is organized as follows.
After preliminary material on automata and learning (\Cref{sec:preliminaries}), in \Cref{sec:incremental} we introduce and prove correctness of our incremental learning algorithm for DFAs, which targets evolving systems whose mutation points are known during learning.
In \Cref{sec:experiments} we benchmark and evaluate the efficiency of the algorithm, and we demonstrate its effectiveness in learning evolving systems.
We conclude in \Cref{sec:discussion} with a discussion of further directions on adaptive automata learning.

\section{Preliminaries}
\label{sec:preliminaries}
Throughout this paper we will make use of standard notation from automata and learning theory. 
We define some of this notation below for the sake of clarity.

We fix a finite alphabet $A$ and write $A^*$ for the set of finite words over this alphabet.
The empty word is $\eword$ and concatenation is written either by juxtaposition or with the operator $\cdot$.
Given a word $w \in A^*$ we write $|w|$ for its length, and $w[i]$, $0 \leq i \leq |w|$, for the prefix of $w$ of length $i$.
A language over $A^*$ is a set of strings such that $L \subseteq A^*$.
We sometimes refer to the characteristic function $A^* \to \mathbb{2} = \{\top, \bot\}$ of $L$ also by $L$ ($L(w) = \top$ if $w \in L$ and $L(w) = \bot$ if $w \notin L$).

The formal models generated by our algorithm are deterministic finite automata (DFAs).
These are 4-tuples $\aut = \langle Q, q_0, \delta, F \rangle$ where $Q$ is a finite set of states, $q_0 \in Q$ is the initial state, $\delta \colon Q \times A \to Q$ is a transition function, and $F \subseteq Q$ is a set of final states.
We define the usual reachability map $\reach_{\aut} \colon A^* \to Q$ by $\reach_{\aut}(\eword) = q_0$ and $\reach_{\aut}(ua) = \delta(\reach_{\aut}(u), a)$.
The language accepted by $\aut$ is $\lang_{\aut} = \{u \in A^* \mid \reach_{\aut}(u) \in F\}$.
We write $A_{\top}$ for the minimal DFA accepting all of $A^*$ and $A_{\bot}$ for the minimal DFA accepting the empty language $\eset$.

Our learner makes use of two types of oracles: \emph{membership} and \emph{equivalence} oracles.
A membership oracle $\mq \colon A^* \to \mathbb{2}$ is able to answer whether a given input word is accepted in the target system; an equivalence oracle takes a DFA $\aut$ and responds with $\eq(\aut) \in A^* \cup \{\mathit{null}\}$, which represents either that the DFA is correct ($\mathit{null}$) or a word $w$ such that $\lang_{\aut}(w) \neq \mq(w)$.

%
%

\subsection{Learning with a Classification Tree}
When designing a learning algorithm, one of the key aspects to consider is how we store the information we acquire over time.
Learning then becomes a matter of being able to extend this structure with as little queries as possible, and transforming the data into a hypothesis automaton. 
%
%
%
The learning algorithm we introduce later, similarly to the classic \kv\ algorithm, uses {\em classification trees} as its base data structure.
%
 Formally, the set of classification trees is given by the following grammar:
\[
    \mathsf{CT} ::= \mathsf{Node}\;A^*\;\mathsf{CT}\;\mathsf{CT}\; \mathop{|}\; \mathsf{Leaf}\;A^*
\]
Here, a node contains a classifier $e \in A^*$, and $\bot$-child and $\top$-child subtrees, and a leaf contains a single access sequence $s \in A^*$.
The child subtrees are named this way because of how the classifier $e$ distinguishes access sequences $s_\bot$ and $s_\top$ present in the respective subtrees: $\mq(s_\bot \cdot e) = \bot$ and $\mq(s_\top \cdot e) = \top$.
In particular, this holds for every pair of leaves and their \textit{lowest common ancestor} node, the root node of the smallest subtree containing both leaves.

The classification tree is then able to classify every word $w \in A^*$ into a specific leaf of the tree, depending on how the target accepts or rejects $w$ concatenated with specific classifiers $e$ in the tree.
This is done through \textit{sifting} (\Cref{alg:sift}), where, starting from the root of the tree with classifier $e$, we pose the query $\mq(w \cdot e)$ and, depending on the result, proceed to the $\bot$-child or the $\top$-child of the node, from which we continue sifting, until we reach a leaf. The access sequence $s$ of that leaf will then be deemed equivalent to $w$.

\begin{subroutine}
	\caption{\texttt{sift} returns the leaf in a classification tree equivalent to a provided word w.r.t.\ the equivalence induced by the tree.}
	\KwData{Classification tree $\mathit{tree}$, membership oracle $\mq$, word $w$.}
	\label[subroutine]{alg:sift}
	\KwResult{Equivalent leaf $l$ in $\mathit{tree}$.}
	$n \gets \mathit{tree}$\;
	\While{$n = \mathsf{Node}\;e\;\mathit{left}\;\mathit{right}$}{
		$n \gets \mq(w \cdot e)\; \texttt{?}\; \mathit{right}\; \texttt{:}\; \mathit{left}$\;
	}
	\Return $n$\;
\end{subroutine}

The leaves of a classification tree represent the discovered states of the hypothesis, and sifting a word down the tree gives us the state this word should reach in the hypothesis.
As such, one can easily retrieve the transitions of the hypothesis from the tree by, for each leaf with access sequence $s$ sifting each extended word $s \cdot a$ to obtain the destination of the transition with symbol $a \in A$ from $s$.
The initial state is simply the state we find by sifting the empty word $\eword$, and the accepting states are the leaves in the $\top$-child subtree of the root node, which will have classifier $\eword$.
This logic is used by \texttt{buildHyp} (\Cref{alg:buildhyp}) to extract the DFA represented by a classification tree.

\begin{subroutine}
	\caption{\texttt{buildHyp} extracts a hypothesis DFA from a classification tree in the Classic KV algorithm.}
	\label[subroutine]{alg:buildhyp}
	\KwData{Classification tree $\mathit{tree}$}
	\KwResult{Updated $\mathcal{H}$ w.r.t the current tree $\mathit{tree}$.}
	$q_o \gets \mathit{sift}(\mathit{tree}, \mq, \varepsilon);\; Q \gets \{q_0\};\; F \gets \varnothing$\;
\For{$l \in \mathit{leaves}(\mathit{tree})$}{
	$Q \gets Q \cup \{l\}$\;
	\If{$l \in \mathit{leaves}(\mathit{child}(\mathit{tree}, \top))$}{
		$F \gets F \cup \{l\}$\;
	}
}

\For{$l \in Q$}{
	\For{$a \in \Sigma$}{
		$\delta(l, a) \gets \mathit{sift}(\mathit{tree}, \mq, \mathit{label}(l) \cdot a)$\;
	}
}

\Return $\langle \Sigma, Q, q_0, F, \delta \rangle$\;	
\end{subroutine}

With a hypothesis extracted from the classification tree, we can now pass this to an equivalence oracle to determine if the hypothesis is correct. If not, we will receive a counterexample word that we can use to improve the classification tree. The algorithm does this by understanding that, given that every hypothesis classifies the empty string $\varepsilon$ correctly, and by definition the current hypothesis classifies the counterexample $c$ incorrectly, there must be a prefix of $c$ for which the classification first diverges. In terms of states, this then means that we are taking a transition into a state that is accepting in the hypothesis, and rejecting in the target, or vice-versa. This is fixed by realising that the state we take this transition from then must actually be two different states. The algorithm then uses this logic in \texttt{updateTree} (\Cref{alg:updatetree}) to split the state into two at the tree level, turning a leaf into a node with two leaves, one representing the new state discovered by the counterexample.

\begin{subroutine}[H]
	\caption{\texttt{updateTree} with a provided counterexample.}
	\label[subroutine]{alg:updatetree}
	\KwData{Classification tree $\mathit{tree}$, counterexample $c \in A^*$.}
	\KwResult{Updated $\mathit{tree}$ taking into account $\mathit{c}$.}
	
	\For{$i \in [0 \cdots \mathit{length}(c) - 1]$}{
		$s_i \gets \mathit{sift}(\mathit{tree}, \mq, c[i]);\;\hat s_i \gets \texttt{reach}_{\mathcal{H}}(c[i])$\;
		\If{$s_i \neq \hat s_i$}{
			$e \gets c_i \cdot \mathit{LCA}(\mathit{tree}, s_i, \hat s_i)$\;
			$\mathit{tree} \gets \mathit{split}(\mathit{tree}, s_{i-1}, c[i-1], e, \mq(c[i-1] \cdot e))$\;
			\Return $\mathit{tree}$\;
		}
	}	
\end{subroutine}



We provide the classic \kv\ algorithm in \Cref{alg:classickv}.
This algorithm uses the \texttt{buildHyp} routine explained above to build a DFA from a classification tree, as well as \texttt{updateTree} (\Cref{alg:updatetree}) to extend the classification tree on receipt of a counterexample from an equivalence oracle.

%
%
%

\begin{algorithm}
	    \caption{Classic Kearns--Vazirani Algorithm}
	    \label[algorithm]{alg:classickv}
    \KwData{Alphabet $A$, membership oracle $\mq$ and equiv. oracle $\eq$ for language $L$.}
	\KwResult{The learned DFA $\mathcal{H}$ accepting $L$.}
	$\mathit{init} \gets \mq(\varepsilon);\; \mathcal{H} \gets \mathit{init}\; \texttt{?}\; A_{\top}\; \texttt{:}\; A_{\bot};\; s \gets \eq(\mathcal{H})$\;
	\If{$s \neq \mathit{null}$}{
		$\mathit{tree} \gets \mathit{init}\; \texttt{?}\; \mathsf{Node}\; \varepsilon\; (\mathsf{Leaf}\; s)\; (\mathsf{Leaf}\; \varepsilon)\; \texttt{:}\; \mathsf{Node}\; \varepsilon\; (\mathsf{Leaf}\; \varepsilon)\; (\mathsf{Leaf}\; s)\;$\;
		$\mathcal{H} \gets \mathtt{buildHyp}(\mathit{tree})$\;
		$\mathit{cex} \gets \eq(\mathcal{H})$\;
		\While{$cex \neq \mathit{null}$}{
			$\mathit{tree} \gets \mathtt{updateTree}(\mathit{tree}, \mathit{cex})$\;
			$\mathcal{H} \gets \mathtt{buildHyp}(\mathit{tree})$\;
			$\mathit{cex} \gets \eq(\mathcal{H})$\;
		}
	}
	\Return $\mathcal{H}$\;
\end{algorithm}

\section{Learning Evolving Systems Incrementally}
\label{sec:incremental}

We now develop a learning algorithm that is able to learn updates to a previous model without having to discard all behavior learned so far, and is also able to detect and remove  behavior that no longer holds.


Classic algorithms are partially able to do this with equivalence oracles---they correct the current hypothesis based on a counterexample. 
{\em Adaptive model learners} are able to do this with the answers of both membership and equivalence oracles, even if the answer conflicts previous ones.
Thus, they can deal with languages that mutate over time, adapting to changes by either trimming outdated behavior or distinguishing new behavior.

Our adaptive learning algorithm is targeted at systems with discrete changes, such as version controlled systems.
Specifically, the system evolves at discrete known points, such as every version, forming a stream of target systems. 
 Our \textit{incremental learning algorithm} for DFAs, presented in \Cref{alg:incremental}, is based on \kv. 
 As a first crucial difference, the incremental algorithm uses a previous learned model as its starting point. As such, it cannot just acquire new information; it also needs to be able to trim outdated behavior. 
This is done by \texttt{minimizeTree} (\Cref{alg:minimizetree}), which prunes an initial tree by removing all leaves that are no longer represented by their reported access sequence. It achieves this by sifting every access sequence down the tree, removing leaves whose access sequences do not sift back into themselves.

\SetKw{Break}{break}
\begin{subroutine}[h]
	\caption{\texttt{minimizeTree} trims the tree from redundant leaves.}
	\label[subroutine]{alg:minimizetree}
	\KwData{Classification tree $\mathit{tree}$, membership oracle $\mq$.}
	\KwResult{A minimized classification tree.}
	\For{$l \in \texttt{leaves}(\mathit{tree})$}{
		$s \gets \texttt{sift}(\mathit{tree}, \mq, \texttt{label}(l))$\;
		\If{$s \neq l$}{
			$\mathit{tree} \gets \texttt{removeLeaf$(\mathit{tree}, l)$}$\;
		}
	}
	\Return $\mathit{tree}$\;
\end{subroutine}

This guarantees not only that the leaves left in our tree are correct in \textit{a} correct automaton for this language, but that every leaf in the tree is unique w.r.t. the Myhill--Nerode congruence. If a pair of leaves were equivalent, then both their access sequences would sift into only one of the nodes, leaving a leaf whose access sequence does not sift back into itself, and causing it to be removed.

\begin{algorithm}[h]
	\caption{Incremental Algorithm}
    \KwData{Fixed alphabet $A$, optional previous classification tree $\mathit{tree}$, membership oracle $\mq$ and equivalence oracle $\eq$ w.r.t the language $L$.}
    \label[algorithm]{alg:incremental}
	\KwResult{The learned DFA $\mathcal{H}$ equivalent to the language $L$.}
	$\mathit{init} \gets \mq(\varepsilon);\; \mathcal{H} \gets \mathit{init}\; \texttt{?}\; A_{\top}\; \texttt{:}\; A_{\bot};\;s \gets \eq(\mathcal{H})$\;
	\If{$s \neq \mathit{null}$}{
		\eIf{$\mathit{tree} = \mathit{null}$}{
			$\mathit{tree} \gets \mathit{init}\; \texttt{?}\; \mathsf{Node}\; \varepsilon\; (\mathsf{Leaf}\; s)\; (\mathsf{Leaf}\; \varepsilon)\; \texttt{:}\; \mathsf{Node}\; \varepsilon\; (\mathsf{Leaf}\; \varepsilon)\; (\mathsf{Leaf}\; s)$\;
		}{
			$\mathit{tree} \gets \texttt{minimizeTree}(\mathit{tree}, \mq)$\;
		}
		$\mathcal{H} \gets \texttt{buildHyp}(\mathit{tree})$\;
		$\mathit{cex} \gets \eq(\mathcal{H})$\;
		\While{$\mathit{cex} \neq \mathit{null}$}{
			$\mathit{tree} \gets \texttt{updateTree}(\mathit{tree}, \mathit{cex})$\;
			$\mathcal{H} \gets \texttt{buildHyp}(\mathit{tree})$\;
			$\mathit{cex} \gets \eq(\mathcal{H})$\;
		}
	}
	\Return $\mathcal{H}$\;
\end{algorithm}

While here we only present the relevant changes made to the classic \kv\ algorithm to be able to adapt to changing behavior, we include the full algorithm with all its subroutines in \Cref{sec:incapp}.

\subsection{Correctness and Termination}
As the algorithm only terminates with a hypothesis that is correct according to an equivalence query, correctness follows from termination. 
Termination of the original \kv\ algorithm relies on the following key property: for every subtree of the form $\mathsf{Node}\;e\;\mathit{left}\;\mathit{right}$, each leaf $s \in A^*$ of $\mathit{left}$ satisfies $se \not\in \lang$ and every leaf $s \in A^*$ of $\mathit{right}$ satisfies $se \in \lang$.
We note that this property also holds in our incremental algorithm as soon as we enter the main loop, as $\mathtt{minimizeTree}$ removes any leaf violating it.

When a counterexample $\mathit{cex}$ is found, the procedure is the same as for the original \kv\ algorithm, and can only be applied a finite number of times: By the property shown above every pair of leaves corresponds to a pair of distinct equivalence classes of the Myhill--Nerode congruence for $\lang$, and therefore the leaves in the tree cannot exceed the number of equivalence classes.
Furthermore, every counterexample of length at least 2 leads to an increase of the number of leaves (via $\mathtt{updateTree}$, which preserves the above invariant).


\section{Experiments}
\label{sec:experiments}

We evaluate the efficiency of our new learning algorithm by running experiments over random targets with different types of features.
While we would like to evaluate it over a standard set of benchmarks~\cite{neider2019}, these currently only cover single target automata, and so are not fit for adaptive learners designed to learn automata that are \emph{linked} due to small evolutions in their behavior.
We designed two scenarios to benchmark this incremental algorithm.
The first scenario takes an initial automaton and applies a series of random mutations: it randomly adds a state, removes a state, diverts a transition, and flips the acceptance of a state.
Our second scenario simulates the common occurrence of adding a feature to an existing system.
We do this by introducing a small \emph{feature automaton} of 3 states to an original base automaton by diverting 3 random transitions into the start state of the feature automaton.

We perform these benchmarks on different automata of increasing number of states, while maintaining the number of mutations applied to them, and the size of the feature automaton. This way, we create different ratios of change, and simulate applying fixes, or adding features to different systems.

We call the first and second targets $t_0$ and $t_1$, respectively.
For our adaptive learning algorithm, the target evolving system starts as $t_0$ and mutates to $t_1$ after 10000 queries.
For the classic \kv\ algorithm, as it cannot learn evolving systems at all, we have to run the algorithm twice, first targeting $t_0$, then from scratch targeting $t_1$.
To ensure repeatable results, each benchmark in question has been run 300 times, each with fresh random inputs of the same parameters, and averaged.
The graphs below represent the average run of both benchmarks, using that both have very similar results.
Separate graphs per benchmark can be found in \Cref{sec:graphsapp}.

We start by running the benchmark on the classic \kv\ algorithm to set a baseline.
These results can be seen in \Cref{fig:mutclassic}.
The progress of each instance here is measured according to the following definition.

\begin{definition}
    Given $\alpha \in [0, 1]$, a stream $(t_i)_{i \in \N}$ of target automata, and a stream $(h_i)_{i \in \N}$ of hypothesis automata,\footnote{Finite streams may be turned into infinite ones by repeating the last element.} the \emph{progress} is the stream $(p_i)_{i \in \N}$, with $p_i \in [0, 1]$ for all $i \in \N$, given by
    \[
        p_i = \sum_{u \in A^*, \lang_{t_i}(u) = \lang_{h_i}(u)} (1 - \alpha) \cdot \left(\frac{\alpha}{|A|}\right)^{|u|}.
    \]
\end{definition}

\begin{figure}[t]
\centering
\begin{minipage}{.45\textwidth}
  \centering
    \resizebox{\textwidth}{!}{%
    	\begin{tikzpicture}
			\begin{axis}[
					xlabel={Number of Queries},
				    ylabel={progress ($\alpha = 0.999$)},
				    ymin=0.5,ymax=1,
			    	xmin=0,xmax=20000,
				    legend pos=south east,
				    legend entries={$|Q|=10$,$|Q|=20$,$|Q|=40$,$|Q|=80$}]
    	  		\addplot[purple] table {./EXP/MIX/MIX-10/classic.dat};
      			\addplot[black] table {./EXP/MIX/MIX-20/classic.dat};
	      		\addplot[blue] table {./EXP/MIX/MIX-40/classic.dat};
    	  		\addplot[orange] table {./EXP/MIX/MIX-80/classic.dat};
			\end{axis}
		\end{tikzpicture}
    }
  \captionof{figure}{Average progress graph of the classic \kv\ algorithm.}
  \label{fig:mutclassic}
\end{minipage}\hfil
\begin{minipage}{.45\textwidth}
  \centering
  	\resizebox{\textwidth}{!}{%
	\begin{tikzpicture}
		\begin{axis}[
				xlabel={Number of Queries},
			    ylabel={progress ($\alpha = 0.999$)},
			    ymin=0.5,ymax=1,
			    xmin=0,xmax=20000,
			    legend pos=south east,
			    legend entries={$|Q|=10$,$|Q|=20$,$|Q|=40$,$|Q|=80$}]
      		\addplot[purple] table {./EXP/MIX/MIX-10/incremental.dat};
      		\addplot[black] table {./EXP/MIX/MIX-20/incremental.dat};
      		\addplot[blue] table {./EXP/MIX/MIX-40/incremental.dat};
      		\addplot[orange] table {./EXP/MIX/MIX-80/incremental.dat};
		\end{axis}
	\end{tikzpicture}
	}
  \captionof{figure}{Average progress graph of the incremental algorithm.}
  \label{fig:mutincremental}
\end{minipage}
\end{figure}

As the learning curves show, the linear process of the classic algorithm for $t_1$ is very similar to the one for $t_0$.
The two halves of the lines may not have the exact same gradients due to the targets being different, but we can see they follow a similar pattern, and more importantly, converge to 1.0 using a similar number of queries.
This is because no knowledge is reused, and all states, even persisting ones, will have to be relearned.

We perform the same benchmark on the incremental algorithm: see results in \Cref{fig:mutincremental}.
The incremental algorithm has a $t_0$ run very similar to the classic algorithm, due to the lack of previous knowledge.
However, the $t_1$ segment is already very different.
We immediately see that it does not start from such a low similarity value as the run of the classic algorithm.
This is because while the classic algorithm always starts from a one state automaton, our incremental algorithm starts from the previous hypothesis, with outdated states pruned.

We can also see that it is not always the case that this line ascends immediately at the mutation point.
This is because, as we are not starting from a widely dissimilar automaton, the equivalence oracle actually requires a number of queries to find a counterexample.
This can be optimized by using more efficient equivalence oracles, but for the sake of simplicity and comparison all algorithms use the same random word search algorithm for equivalence testing.

\begin{figure}[t]
\centering
\begin{minipage}{.5\textwidth}
	\centering
	\resizebox{\textwidth}{!}{%
  	\begin{tikzpicture}
		\begin{axis}[
				xlabel={State Space ($|Q|$)},
			    ylabel=Average Ratio,
			    ymin=0.35,ymax=1.3,
			    xmin=0,xmax=170,
			    legend pos=north east,
			    cycle list name=black white,
			    legend entries={Baseline,Mutation (Classic), Mutation (Adaptive), Feature-Add (Classic), Feature-Add (Adaptive)}]
			    \addplot[mark=none, dashed] coordinates {(0,1) (170,1)};
      		\addplot[smooth,mark=square*, purple] table {./EXP/mut.dat};
      		\addplot[smooth,mark=*, black] table[y index=2] {./EXP/mut.dat};
      		\addplot[smooth,mark=square*, blue] table {./EXP/feat.dat};
      		\addplot[smooth,mark=*, orange] table[y index=2] {./EXP/feat.dat};
		\end{axis}
	\end{tikzpicture}
	}
\end{minipage}\hfil
  \caption{Ratio of number of queries needed to learn the automata.}
  \label{fig:mutratio}
\end{figure}
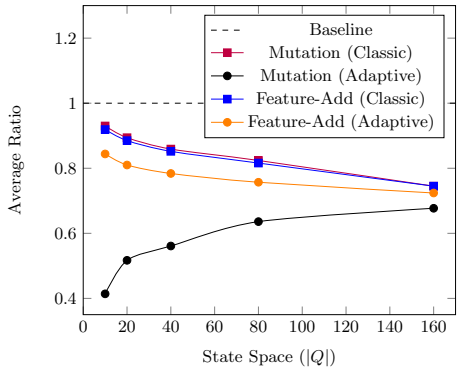

These learning progress graph representations are great at demonstrating the overall behavior and approach of the learning algorithm, but we now want to evaluate whether this algorithm does indeed provide a benefit over learning systems classically, or with the current state-of-the-art adaptive algorithms, in terms of the number of queries it takes them to fully learn mutated targets.
As such, we have computed and averaged the number of queries it takes to reach the final hypothesis while learning the $t_1$ on each run.
We present the results as two ratios relative to the incremental algorithm: one comparing the classic \kv\ algorithm, and another comparing the current state-of-the-art adaptive algorithm, Partial Dynamic \lstar~\cite{damasceno2019}.
We plot these in \Cref{fig:mutratio}, and show how the ratio changes with the size of the state space of $t_0$.

In this representation, a value of 1.0 would indicate that the incremental algorithm requires as many queries as the algorithm in comparison. Any value below indicates a benefit in running the incremental algorithm over such algorithm, and vice versa (the lower the ratio, the better the incremental algorithm).

As we can see, {\bf the incremental algorithm consistently outperforms all other algorithms in all benchmarks}, even with small targets where a relatively big portion of the system has changed.
When compared to the adaptive algorithm from the literature, we can see the incremental algorithm still consistently outperforms in terms of queries, with a tendency to plateau between a 0.75 and 0.7 ratio---a reduction by 25--30\% in the number of queries.

\section{Conclusion}
\label{sec:discussion}
We introduced a new state-of-the-art algorithm for \emph{adaptive learning} which provides, to our knowledge, the most efficient adaptive learner to date, allowing us to learn systems that were before too big or evolved too quickly to be learned classically. We evaluated the fitness of this algorithm through a set of realistic experiments, demonstrating their benefit over classic and existing adaptive learners. Adaptive learning could provide a cooperative relationship between active and passive learning, due to its flexibility towards information that changes over time and becomes available at different stages. We want to explore this in future work, as well as developing a formal framework of adaptive learning, where other algorithms can be easily adapted, e.g. efficient classic algorithms such as TTT~\cite{isberner2015}. Finally, although our incremental algorithm allows us to learn systems that evolve at known points, this would not work in a true black-box scenario, where we cannot know if or when the system changes. In the future we would like to then develop a \emph{continuous} learning algorithm for such evolving systems.

\paragraph{\bf Related Work}
Our algorithm contributes to the field of adaptive learning first introduced in~\cite{groce2002}, where information learned in previous models was used as guidance for which states to check first, instead of blindly looking for new ones. This was done by slightly modifying the \lstar\ algorithm to start from a previous set of access sequences. Chaki et al.~\cite{chaki2008} use a similar algorithm in combination with assume-guarantee reasoning~\cite{jones1983} to provide a framework where model checking is used to find counterexamples in the model, and thus make progress in learning. Their algorithm, Dynamic \lstar, reuses not only the starting prefix/suffix sets, but also their computed values. However, these must still be validated on the new target. Finally, Damasceno et al.~\cite{damasceno2019} introduce Partial Dynamic \lstar, which improves Dynamic \lstar\ by analysing the start prefix/suffix sets to trim them where possible, reducing the amount of information that needs to be validated.

These previous algorithms, however, suffer from using an observation table as their data structure, which increases the amount of redundant data to be acquired. As the tables grow, these redundancies significantly increase the number of queries that need to be asked.


\FloatBarrier
\bibliographystyle{splncs04}
\bibliography{keep-learning}

\begin{thebibliography}{10}
\providecommand{\url}[1]{\texttt{#1}}
\providecommand{\urlprefix}{URL }
\providecommand{\doi}[1]{https://doi.org/#1}

\bibitem{angluin1987}
Angluin, D.: Learning regular sets from queries and counterexamples.
  Information and Computation  \textbf{75},  87--106 (1987).
  \doi{10.1016/0890-5401(87)90052-6}

\bibitem{bornholt2021}
Bornholt, J., Joshi, R., Astrauskas, V., Cully, B., Kragl, B., Markle, S.,
  Sauri, K., Schleit, D., Slatton, G., Tasiran, S., Van~Geffen, J., Warfield,
  A.: Using {Lightweight} {Formal} {Methods} to {Validate} a {Key}-{Value}
  {Storage} {Node} in {Amazon} {S3}. In: SIGOPS. pp. 836--850. ACM (2021).
  \doi{10.1145/3477132.3483540}

\bibitem{chaki2008}
Chaki, S., Clarke, E.M., Sharygina, N., Sinha, N.: Verification of evolving
  software via component substitutability analysis. Formal Methods Syst. Des.
  \textbf{32}(3),  235--266 (2008). \doi{10.1007/s10703-008-0053-x}

\bibitem{damasceno2019}
Damasceno, C.D.N., Mousavi, M.R., da~Silva~Sim{\~{a}}o, A.: Learning to reuse:
  Adaptive model learning for evolving systems. In: IFM. LNCS, vol. 11918, pp.
  138--156. Springer (2019). \doi{10.1007/978-3-030-34968-4\_8}

\bibitem{FerreiraBD021}
Ferreira, T., Brewton, H., D'Antoni, L., Silva, A.: Prognosis: closed-box
  analysis of network protocol implementations. In: SIGCOMM. pp. 762--774.
  {ACM} (2021). \doi{10.1145/3452296.3472938}

\bibitem{Fiterau-Brostean16}
Fiterau{-}Brostean, P., Janssen, R., Vaandrager, F.W.: Combining model learning
  and model checking to analyze {TCP} implementations. In: CAV. LNCS,
  vol.~9780, pp. 454--471. Springer (2016). \doi{10.1007/978-3-319-41540-6\_25}

\bibitem{Fiterau-Brostean17}
Fiterau{-}Brostean, P., Lenaerts, T., Poll, E., de~Ruiter, J., Vaandrager,
  F.W., Verleg, P.: Model learning and model checking of {SSH} implementations.
  In: SPIN. pp. 142--151. {ACM} (2017). \doi{10.1145/3092282.3092289}

\bibitem{groce2002}
Groce, A., Peled, D.A., Yannakakis, M.: Adaptive model checking. In: TACAS.
  LNCS, vol.~2280, pp. 357--370. Springer (2002).
  \doi{10.1007/3-540-46002-0\_25}

\bibitem{huistra2018}
Huistra, D., Meijer, J., van~de Pol, J.: Adaptive learning for learn-based
  regression testing. In: FMICS. LNCS, vol. 11119, pp. 162--177. Springer
  (2018). \doi{10.1007/978-3-030-00244-2\_11}

\bibitem{ttt2014}
Isberner, M., Howar, F., Steffen, B.: The {TTT} {Algorithm}: {A}
  {Redundancy}-{Free} {Approach} to {Active} {Automata} {Learning}. In: Runtime
  {Verification}, LNCS, vol.~8734, pp. 307--322. Springer International
  Publishing, Cham (2014). \doi{10.1007/978-3-319-11164-3\_26}

\bibitem{isberner2015}
Isberner, M., Howar, F., Steffen, B.: The open-source {L}earn{L}ib. In: CAV.
  LNCS, vol.~9206, pp. 487--495 (2015). \doi{10.1007/978-3-319-21690-4\_32}

\bibitem{jones1983}
Jones, C.B.: Tentative steps toward a development method for interfering
  programs. ACM Transactions on Programming Languages and Systems  \textbf{5},
  596--619 (1983). \doi{10.1145/69575.69577}

\bibitem{kv1994}
Kearns, M.J., Vazirani, U.V.: An introduction to computational learning theory.
  MIT Press, Cambridge, Mass (1994),
  \url{https://mitpress.mit.edu/books/introduction-computational-learning-theory}

\bibitem{neider2019}
Neider, D., Smetsers, R., Vaandrager, F., Kuppens, H.: Benchmarks for
  {Automata} {Learning} and {Conformance} {Testing}. In: Margaria, T., Graf,
  S., Larsen, K.G. (eds.) Models, {Mindsets}, {Meta}: {The} {What}, the {How},
  and the {Why} {Not}? {Essays} {Dedicated} to {Bernhard} {Steffen} on the
  {Occasion} of {His} 60th {Birthday}, pp. 390--416. Springer International
  Publishing, Cham (2019). \doi{10.1007/978-3-030-22348-9\_23},
  \url{https://doi.org/10.1007/978-3-030-22348-9_23}

\bibitem{frits-cacm}
Vaandrager, F.W.: Model learning. Commun. {ACM}  \textbf{60},  86--95 (2017).
  \doi{10.1145/2967606}

\bibitem{windmuller2013}
Windm{\"{u}}ller, S., Neubauer, J., Steffen, B., Howar, F., Bauer, O.: Active
  continuous quality control. In: CBSE. pp. 111--120. {ACM} (2013).
  \doi{10.1145/2465449.2465469}

\end{thebibliography}

\FloatBarrier
\clearpage

\appendix

\setcounter{theorem}{0}
\renewcommand{\thetheorem}{A.\arabic{theorem}}
\setcounter{algocf}{0}
\renewcommand{\thealgocf}{A.\arabic{algocf}}

\section{Omitted Incremental Subroutines}
\label{sec:incapp}

\begin{subroutine}
	\caption{\texttt{child} returns the ($\bot$/$\top$)-child of a provided tree node.}
	\label[subroutine]{alg:child}
	\KwData{Tree node $p$, $\mathit{b} \in \mathbb{2}$}
	\KwResult{$\mathit{b}$-child of $p$.}
	\If{$p = \mathsf{Node}\;e\;\mathit{left}\;\mathit{right}$}{
		\Return $\mathit{b}\; \texttt{?}\; \mathit{right}\; \texttt{:}\; \mathit{left}$\;
	}
	\Return $\mathit{null}$\;
\end{subroutine}

\begin{subroutine}
	\caption{\texttt{children} returns the children of a provided tree node.}
	\KwData{Tree node $p$.}
	\KwResult{Set of children of $p$.}
	$C \gets \varnothing$\;
	\If{$\mathit{child}(p, \bot) \neq \mathit{null}$}{
		$C \gets C \cup \{ \mathit{child}(p, \bot) \}$\;
	}
	\If{$\mathit{child}(p, \top) \neq \mathit{null}$}{
		$C \gets C \cup \{ \mathit{child}(p, \top) \}$\;
	}
	\Return $C$\;
\end{subroutine}

\begin{subroutine}
	\caption{\texttt{setChild} updates a child of a given inner node in a classification tree.}
	\KwData{Classification tree $\mathit{tree}$, parent node $p$, child outcome $b$, new child $n$.}
	\label[subroutine]{alg:setchild}
	\KwResult{Updated classification tree $\mathit{tree}$ where $n$ is the $b$-child of $p$.}
	\If{$\mathit{tree} = \mathsf{Node}\;e\;\mathit{left}\;\mathit{right}$}{
		\If{$\mathit{tree} = p$}{
			\eIf{$b$}{
				\Return $\mathsf{Node}\;e\;\mathit{left}\;\mathit{n}$\;
			}{
				\Return $\mathsf{Node}\;e\;\mathit{n}\;\mathit{right}$\;
			}
		}
		\Return $\mathsf{Node}\; e\; \mathit{setChild}(\mathit{left}, p, b, n)\; \mathit{setChild}(\mathit{right}, p, b, n)$\;
		
	}
	\Return $\mathit{tree}$\;
\end{subroutine}

\begin{subroutine}
	\caption{\texttt{nodes} returns the set of all nodes in a given tree.}
	\KwData{Classification tree $\mathit{tree}$.}
	\KwResult{Set of nodes in $\mathit{tree}$.}
	\If{$\mathit{tree} = \mathsf{Leaf}\; s$}{
		\Return $\{\mathit{tree}\}$\;
	}
	\Return $\mathit{tree} \cup \mathit{nodes}(\mathit{child}(\mathit{tree}, \bot)) \cup \mathit{nodes}(\mathit{child}(\mathit{tree}, \top))$\;
\end{subroutine}

\begin{subroutine}
	\caption{\texttt{leaves} returns the set of leaves in a given tree.}
	\KwData{Classification tree $\mathit{tree}$.}
	\KwResult{Set of leaves in $\mathit{tree}$.}
	\If{$\mathit{tree} = \mathsf{Leaf}\; s$}{
		\Return $\{\mathit{tree}\}$\;
	}
	\Return $\mathit{leaves}(\mathit{child}(\mathit{tree}, \bot)) \cup \mathit{leaves}(\mathit{child}(\mathit{tree}, \top))$\;
\end{subroutine}

\begin{subroutine}
	\caption{\texttt{label} returns the label of a given node, be it a classifier or an access sequence.}
	\KwData{Node $n$.}
	\KwResult{Label $\in A^*$.}
	\If{$n = \mathsf{Leaf}\; s$}{
		\Return $s$\;
	}
	\If{$n = \mathsf{Node}\;e\;\mathit{left}\;\mathit{right}$}{
		\Return $e$\;
	}
\end{subroutine}

\begin{subroutine}
	\caption{\texttt{setLabel} replaces the label in a specific leaf.}
	\KwData{Classification tree $\mathit{tree}$, leaf $l$, new label $w$.}
	\KwResult{Updated classification tree $\mathit{tree}$ with new label $w$ in $l$.}
	\If{$\mathit{tree} = l$}{
		\Return $\mathsf{Leaf}\; w$\;
	}
	\If{$\mathit{tree} = \mathsf{Node}\;e\;\mathit{left}\;\mathit{right}$}{
		\Return $\mathsf{Node}\; e\; \mathit{setLabel}(\mathit{left}, l, w)\;
									\mathit{setLabel}(\mathit{right}, l, w)$\;
	}
	\Return $\mathit{tree}$\;
\end{subroutine}

\begin{subroutine}
	\caption{\texttt{outcome} returns whether a provided node $n$ is a $\bot$-child or a $\top$-child, or neither.}
	\KwData{Classification tree $\mathit{tree}$, provided node $n$}
	\label[subroutine]{alg:outcome}
	\KwResult{$\bot$, $\top$ or $\mathit{null}$ if the node does not have a parent in $\mathit{tree}$}
	\If{$n = \mathit{tree}$}{
		\Return $\mathit{null}$\;
	}
	\Return $\mathit{child}(\mathit{parent}(\mathit{tree}, n), \top) = n$\;
	
\end{subroutine}

\begin{subroutine}
	\caption{\texttt{parent} returns the parent of a provided node in the classification tree.}
	\KwData{Classification tree $\mathit{tree}$, child leaf $l$.}
	\label[subroutine]{alg:parent}
	\KwResult{Parent node $p$.}
	$q \gets \mathit{Queue}(\mathit{tree})$\;
	\While{$|q| \neq 0$}{
		$p \gets \mathit{pop}(q)$\;
		\If{$l \in \mathit{children}(p)$}{
			\Return $p$\;
		}
		\For{$c \in \mathit{children}(p)$}{
			$q \gets \mathit{push}(q, c)$\;			
		}
	}
	\Return $\mathit{null}$\;
\end{subroutine}

\begin{subroutine}
	\caption{\texttt{removeLeaf}}
	\label[subroutine]{alg:removeleaf}
	\KwData{Classification tree $\mathit{tree}$, leaf $l \in \mathtt{leaves}(\mathit{tree})$ to be removed.}
	\KwResult{A valid classification tree $\mathit{tree}$ with $l$ removed.} 
	$\mathit{node} \gets \texttt{parent}(\mathit{tree}, l)$\;
	$\mathit{sibling} \gets \texttt{child}(\mathit{node}, \neg\texttt{outcome}(\mathit{tree}, l))$\;
	\Return $\texttt{setChild}(\mathit{tree}, \texttt{parent}(\mathit{tree}, \mathit{node}), \texttt{outcome}(\mathit{tree}, \mathit{node}), \mathit{sibling})$\;
\end{subroutine}

\begin{subroutine}
	\caption{\texttt{LCA} returns the lowest common ancestor node of two provided leaves in a classification tree.}
	\KwData{Classification tree $\mathit{tree}$, first leaf $l_a$, second leaf $l_b$.}
	\KwResult{LCA node $n$ in the classification tree $\mathit{tree}$.}
	$n \gets l_a$\;
	\While{$l_b \not \in \mathit{leaves}(n)$}{
		$n \gets \mathit{parent}(\mathit{tree}, n)$\;
	}
	\Return $n$\;
\end{subroutine}

\begin{subroutine}
	\caption{\texttt{split} splits a leaf in the tree into a node with 2 child leaves, one of them a new leaf introduced to the tree.}
	\KwData{Classification tree $\mathit{tree}$, current leaf $l$ being split, label $w \in A^*$ for the new leaf, classifier $e \in A^*$ for the new node, $b \in \mathbb{2}$ indicating whether the new leaf is a $\top$-child.}
	\KwResult{Updated classification tree $\mathit{tree}$.}
	$\mathit{node} \gets b\;
		\texttt{?}\; \mathsf{Node}\; e\; l\; (\mathsf{Leaf}\;w)\;
		\texttt{:}\; \mathsf{Node}\; e\; (\mathsf{Leaf}\;w)\; l$\;
	\Return $\mathit{setChild}(\mathit{tree}, \mathit{parent}(\mathit{tree}, l), \mathit{outcome}(\mathit{tree}, l), \mathit{node})$\;
\end{subroutine}

\FloatBarrier
\section{Additional Experiment Graphs}
\label{sec:graphsapp}

\FloatBarrier
\subsection{Mutation Benchmark}

\begin{figure}
	\centering
	\resizebox{0.62\columnwidth}{!}{%
		\begin{tikzpicture}
			\begin{axis}[
				xlabel={Number of Queries},
			    ylabel={progress ($\alpha = 0.999$)},
			    ymin=0.5,ymax=1,
			    xmin=0,xmax=20000,
			    legend pos=south east,
			    legend entries={$|Q|=10$,$|Q|=20$,$|Q|=40$,$|Q|=80$}]
	      		\addplot[purple] table {./EXP/MUT/MUT-10/classic.dat};
	      		\addplot[black] table {./EXP/MUT/MUT-20/classic.dat};
	      		\addplot[blue] table {./EXP/MUT/MUT-40/classic.dat};
	      		\addplot[orange] table {./EXP/MUT/MUT-80/classic.dat};
			\end{axis}
		\end{tikzpicture}
	}
	\caption{Average progress graph of the classic \kv\ algorithm.}
\end{figure}
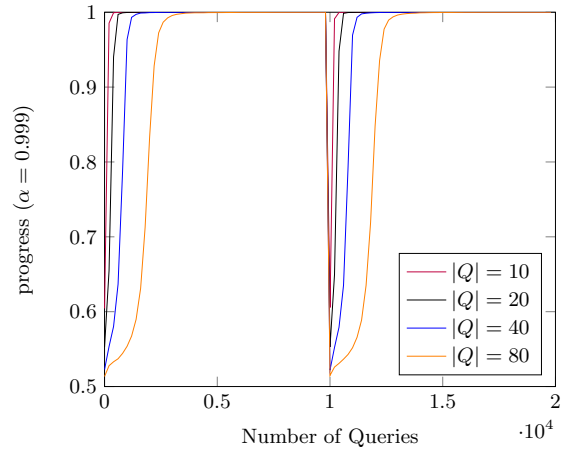

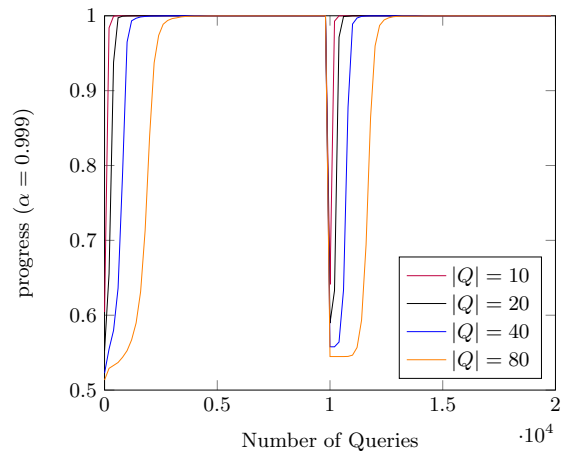
\begin{figure}
	\centering
	\resizebox{0.62\columnwidth}{!}{%
		\begin{tikzpicture}
			\begin{axis}[
				xlabel={Number of Queries},
			    ylabel={progress ($\alpha = 0.999$)},
			    ymin=0.5,ymax=1,
			    xmin=0,xmax=20000,
			    legend pos=south east,
			    legend entries={$|Q|=10$,$|Q|=20$,$|Q|=40$,$|Q|=80$}]
	      		\addplot[purple] table {./EXP/MUT/MUT-10/incremental.dat};
	      		\addplot[black] table {./EXP/MUT/MUT-20/incremental.dat};
	      		\addplot[blue] table {./EXP/MUT/MUT-40/incremental.dat};
	      		\addplot[orange] table {./EXP/MUT/MUT-80/incremental.dat};
			\end{axis}
		\end{tikzpicture}
	}
	\caption{Average progress graph of the incremental algorithm.}
\end{figure}

\FloatBarrier
\pagebreak
\subsection{Feature-Add Benchmark}

\begin{figure}
	\centering
	\resizebox{0.62\columnwidth}{!}{%
		\begin{tikzpicture}
			\begin{axis}[
				xlabel={Number of Queries},
			    ylabel={progress ($\alpha = 0.999$)},
			    ymin=0.5,ymax=1,
			    xmin=0,xmax=20000,
			    legend pos=south east,
			    legend entries={$|Q|=10$,$|Q|=20$,$|Q|=40$,$|Q|=80$}]
	      		\addplot[purple] table {./EXP/FEAT/FEAT-10/classic.dat};
	      		\addplot[black] table {./EXP/FEAT/FEAT-20/classic.dat};
	      		\addplot[blue] table {./EXP/FEAT/FEAT-40/classic.dat};
	      		\addplot[orange] table {./EXP/FEAT/FEAT-80/classic.dat};
			\end{axis}
		\end{tikzpicture}
	}
	\caption{Average progress graph of the classic \kv\ algorithm.}
\end{figure}
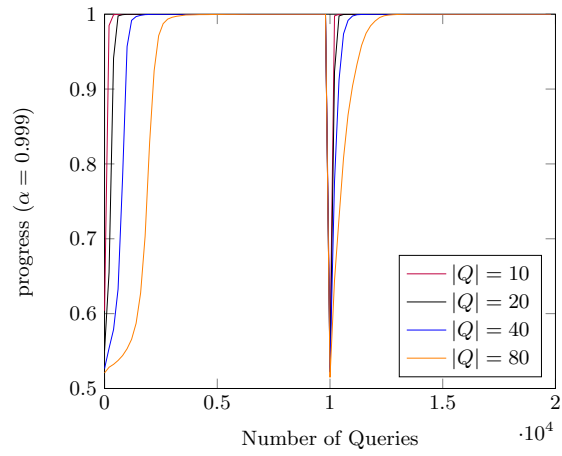

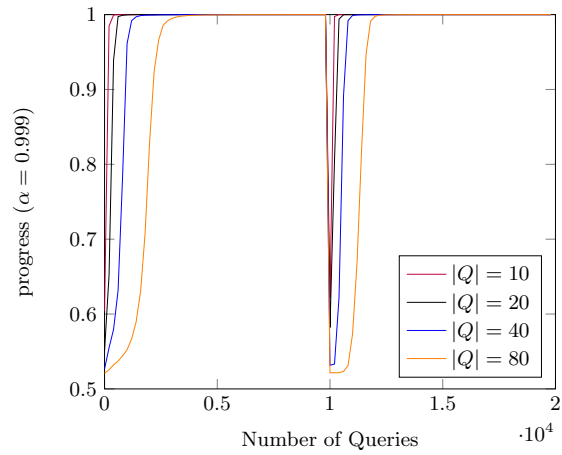
\begin{figure}
	\centering
	\resizebox{0.62\columnwidth}{!}{%
		\begin{tikzpicture}
			\begin{axis}[
				xlabel={Number of Queries},
			    ylabel={progress ($\alpha = 0.999$)},
			    ymin=0.5,ymax=1,
			    xmin=0,xmax=20000,
			    legend pos=south east,
			    legend entries={$|Q|=10$,$|Q|=20$,$|Q|=40$,$|Q|=80$}]
	      		\addplot[purple] table {./EXP/FEAT/FEAT-10/incremental.dat};
	      		\addplot[black] table {./EXP/FEAT/FEAT-20/incremental.dat};
	      		\addplot[blue] table {./EXP/FEAT/FEAT-40/incremental.dat};
	      		\addplot[orange] table {./EXP/FEAT/FEAT-80/incremental.dat};
			\end{axis}
		\end{tikzpicture}
	}
	\caption{Average progress graph of the incremental algorithm.}
\end{figure}

\end{document}